\documentclass[journal]{IEEEtran}

\usepackage{tikz}
\usetikzlibrary{decorations.pathreplacing}
\usetikzlibrary{arrows,shapes,positioning}
\usetikzlibrary{patterns}
\usepackage{pgfplots}
\pgfplotsset{compat=newest}
\pgfplotsset{plot coordinates/math parser=false}
\newlength\figureheight
\newlength\matlabfigurewidth
\usepackage{cite}
\graphicspath{{fig/}}

\usepackage{siunitx}

\usepackage{cite}

\ifCLASSINFOpdf
\else
\fi
\usepackage{amsmath}
\usepackage{algorithmic}
\usepackage{array}
\usepackage{mathrsfs}
\usepackage{mathtools}

\ifCLASSOPTIONcompsoc
  \usepackage[caption=false,font=normalsize,labelfont=sf,textfont=sf]{subfig}
\else
  \usepackage[caption=false,font=footnotesize]{subfig}
\fi
\usepackage{fixltx2e}

\begin{document}
%
\title{Photovoltaic Model-Based Solar Irradiance Estimators: Performance Comparison and  Application to  Maximum Power Forecasting}
%
%
%
\author{Enrica~Scolari,~\IEEEmembership{Member,~IEEE,}
		Fabrizio~Sossan,~\IEEEmembership{Member,~IEEE,}
		and~Mario~Paolone,~\IEEEmembership{Senior~Member,~IEEE}
	\thanks{The Authors are with the Ecole Polytechnique Fédérale de Lausanne EPFL, ´
		CH-1015, Lausanne, Switzerland.}
	\thanks{This research received funding from the Swiss Competence Center for
			Energy Research (FURIES) and by the SNSF NRP70 “Energy Turnaround”.}}
\maketitle

\begin{abstract}
Due to the increasing proportion of distributed photovoltaic (PV) production in the generation mix, the knowledge of the PV generation capacity has become a key factor. In this work, we propose to compute the PV plant maximum power starting from the indirectly-estimated irradiance. Three estimators are compared in terms of i) ability to compute the PV plant maximum power, ii) bandwidth and iii) robustness against measurements noise.  The approaches rely on measurements of the DC voltage, current, and cell temperature and on a model of the PV array. We show that the considered methods can accurately reconstruct the PV maximum generation even during curtailment periods, i.e. when the measured PV  power is not representative of the maximum potential of the PV array. Performance evaluation is carried out by using a dedicated experimental setup on a 14.3 kWp rooftop PV installation.
Results also proved that the analyzed methods can outperform pyranometer-based estimations, with a less complex sensing system. We show how the obtained PV maximum power values can be applied to train time series-based solar maximum power forecasting techniques. This is beneficial when the measured power values, commonly used as training, are not representative of the maximum PV potential.

\end{abstract}


\IEEEdisplaynontitleabstractindextext

%
\section{Introduction}\label{sec:introduction}
{\color{black} \IEEEPARstart{E}{stimating} the photovoltaic (PV) maximum generation from real-time measurements is a key factor to enable the definition of robust control schemes capable to enforce the safe operation of power distribution networks with densely clustered PV plants, e.g. \cite{liu2008distribution,Journal:BernsteinReyes:2015:COMMELEC:Part1}. The knowledge of the real-time  PV generation potential is an important input for self-consumption strategies,\cite{selfconsumption1,selfconsumption2}, local control of power systems, \cite{Hossain2012, nick2014}, and creation of historical datasets. These datasets are widely used for both the training of data-driven forecasting models, \cite{li2009, pinson2016,Inman2013535, review2016}, and to infer local behavior of PV generation starting from information at higher spatial and time resolution, \cite{proxy}. 

The knowledge of the PV generation is normally achieved by directly measuring the power injected into the grid by the PV inverter. However, local measurements might be affected by the action of (unobserved)  control actions,  e.g. due to curtailment policies \cite{curtailment2}.  These exogenous disturbances would make the observations uncorrelated with the true PV maximum power and, consequently, with the true irradiance. Due to their lack of representativeness, the use of these observations as proxy measurements or as training data for PV forecasting would have a detrimental impact on the estimation of the PV state and prediction performance, respectively. 
In this context, we propose  to reconstruct the maximum power production of a PV plant using model-based approaches that need measurements of the array DC voltage, DC current, and cell temperature as inputs. They are based on the single diode five-parameter model proposed by \cite{desoto}, extended to a whole PV plant.

Similar approaches have already been exploited in the existing literature to estimate the irradiance from a single PV module. 
 Authors of \cite{Laudani1} propose a closed-form analytical estimator of the irradiance based on the same inputs considered in this work. Similarly, Authors of \cite{carrasco2014} propose a globally convergent estimator based on the immersion and invariance ($ I\&I $) principle.}
 In \cite{vigni2015}, temperature and DC electrical measurements are used to perform real-time estimation of the irradiance. The method has the drawback of requiring the PV system to move in three different states (panel under load, short circuit or open circuit), a feature that is not normally implemented in commercial PV systems. 
{\color{black}In \cite{Laudani2}, an approach based on neural network is also proposed to estimate the irradiance starting from cell temperature and electrical DC information. The proposed model is implemented in a microcontroller and used to infer the irradiance.}
In \cite{cristaldi2014}, a  model-based approach is used to implement a maximum power point tracking (MPPT) algorithm. The method requires a pyranometer for the identification of the model parameters and is validated for a single module.

{\color{black}It is to note that there are some advantages in estimating the power output using measurements of electrical quantities rather than information from a pyranometer.  On one hand, voltage and current measurements are generally available from the converter management system and the module temperature is easy to measure, \cite{Laudani2}. On the other hand, irradiance sensors are sensitive to calibration, \cite{Myers2005}, and return measurements that are significative for a specific point rather than gathering the average irradiance conditions on the installation, \cite{arena2015}. For this last reason, several pyranometers might be necessary, this increasing the overall cost. 
Furthermore, when observations from a pyranometer are not available nor reliable, one might consider to estimate the irradiance and the PV production potential by exploiting the knowledge of the measured production of nearby PV installations. Also in this case, one should consider that the PV power output could be curtailed, thus the measured power would not be representative of the true irradiance potential and its maximum value should be reconstructed.}

In this paper,  three model-based methods are used to reconstruct the maximum power of a PV plant starting from experimental measurements of the DC voltage, current and temperature.
{\color{black} Two formulations from the existing literature (the analytical from \cite{Laudani1} and the I\&I from \cite{carrasco2014}) are considered and extended to estimate the theoretical maximum power of an entire PV array, irrespectively of its operating conditions.} A third estimator, based on the Kalman Filter (KF) formulation, is proposed here and included in the performance assessment. 
{\color{black} More specifically, the contributions of the paper are listed here below:
\begin{itemize}
	\item  a formal comparison of  PV maximum power estimators is performed using experimental data from a 14.3~kWp PV-roof installation;
	\item  performance is validated  with measurements from a dedicated experimental setup which allows to account for MPPT and non-MPPT operating conditions;
	\item  the maximum power estimations of the analyzed methods are compared with those obtained starting from pyranometer readings;
	\item  the rejection to measurements noise and the bandwidth of the estimators are formally compared. \footnote{If the analytical method is expected to accurately capture the irradiance dynamics, the KF and I$\&$I-based estimations are expected to have improved rejection against noise.}
	\item a practical application of the methods to improve data-driven maximum power forecasting tools is proposed. We show that training the forecasting algorithm with historical data that are corrupted by exogenous disturbances might lead to a deterioration of the forecasting tool, in terms of its ability to predict the maximum power. In this context, we apply the analyzed methods to reconstruct the maximum point and we therefore make sure that the forecasting algorithm learns from values that are always representative of the true irradiance potential.
\end{itemize} }

The paper is organized as follows: Section~\ref{PVmodel} introduces the PV model selected for the analysis.  Section~\ref{Estimators} describes the three methods proposed to estimate the irradiance while Section~\ref{power} explains how we compute the maximum power.  Section~\ref{setup} shows the experimental setup and Section~\ref{results} discusses the main results. Section~\ref{application} explains how the method can be used to improve time series-based power forecast. Section~\ref{conclusion} draws the main conclusions.

\section{The PV single-diode model}\label{PVmodel}
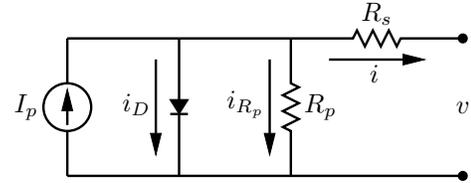
\begin{figure}[h!]
	\centering
	\vspace{-3mm}
	\begin{tikzpicture}[scale=2.54]
\ifx\dpiclw\undefined\newdimen\dpiclw\fi
\global\def\dpicdraw{\draw[line width=\dpiclw]}
\global\def\dpicstop{;}
\dpiclw=0.8bp
\dpiclw=0.8bp
\dpiclw=0.9bp
\dpicdraw (0,0)
 --(0,0.235)\dpicstop
\dpicdraw (0,0.36) circle (0.049213in)\dpicstop
\dpicdraw (0,0.26625)
 --(0,0.34125)\dpicstop
\filldraw[line width=0bp](0.028125,0.34125)
 --(0,0.45375)
 --(-0.028125,0.34125) --cycle
\dpicstop
\dpicdraw (0,0.485)
 --(0,0.72)\dpicstop
\draw (-0.125,0.36) node[left=-1.5bp]{$ I_p$};
\dpicdraw (0,0.72)
 --(0.585,0.72)\dpicstop
\dpicdraw (0.585,0.72)
 --(0.585,0.396084)\dpicstop
\global\let\dpicshdraw=\dpicdraw\global\def\dpicdraw{}
\global\def\dpicstop{--}
\dpicshdraw[fill=white!0!black]
\dpicdraw (0.585,0.396084)
 --(0.626667,0.396084)
 --(0.585,0.330142)
 --(0.543333,0.396084)
 --(0.585,0.396084)\dpicstop
cycle; \global\let\dpicdraw=\dpicshdraw\global\def\dpicstop{;}
\dpicdraw (0.538775,0.323916)
 --(0.631225,0.323916)\dpicstop
\dpicdraw (0.585,0.323916)
 --(0.585,0)\dpicstop
\dpicdraw (0.463363,0.61)
 --(0.463363,0.2225)\dpicstop
\filldraw[line width=0bp](0.435238,0.2225)
 --(0.463363,0.11)
 --(0.491488,0.2225) --cycle
\dpicstop
\draw (0.463363,0.372885) node[left=-1.5bp]{$ i_D$};
\dpicdraw (0.585,0.72)
 --(1.17,0.72)\dpicstop
\dpicdraw (1.17,0.72)
 --(1.17,0.485)
 --(1.211667,0.464167)
 --(1.128333,0.4225)
 --(1.211667,0.380833)
 --(1.128333,0.339167)
 --(1.211667,0.2975)
 --(1.128333,0.255833)
 --(1.17,0.235)
 --(1.17,0)\dpicstop
\draw (1.211667,0.36) node[right=-1.5bp]{$ R_{p}$};
\dpicdraw (1.059148,0.61)
 --(1.059148,0.2225)\dpicstop
\filldraw[line width=0bp](1.031023,0.2225)
 --(1.059148,0.11)
 --(1.087273,0.2225) --cycle
\dpicstop
\draw (1.059148,0.372885) node[left=-1.5bp]{$ i_{R_p}$};
\dpicdraw (1.17,0.72)
 --(1.495,0.72)
 --(1.515833,0.761667)
 --(1.5575,0.678333)
 --(1.599167,0.761667)
 --(1.640833,0.678333)
 --(1.6825,0.761667)
 --(1.724167,0.678333)
 --(1.745,0.72)
 --(2.07,0.72)\dpicstop
\draw (1.62,0.761667) node[above=-1.5bp]{$ R_{s}$};
\dpicdraw (1.37,0.609148)
 --(1.7575,0.609148)\dpicstop
\filldraw[line width=0bp](1.7575,0.581023)
 --(1.87,0.609148)
 --(1.7575,0.637273) --cycle
\dpicstop
\draw (1.607115,0.609148) node[below=-1.5bp]{$ i$};
\dpicdraw[fill=black](2.07,0.72) circle (0.007874in)\dpicstop
\draw (2.07,0.36) node{$v$};
\dpicdraw[fill=black](2.07,0) circle (0.007874in)\dpicstop
\dpicdraw (2.07,0)
 --(0,0)\dpicstop
\end{tikzpicture}
	\caption{The adopted five parameter circuit model of a PV cell.}
	\label{Figure:PVcirc}
\end{figure}

The estimators described in Section~ \ref{Estimators} all rely on a physical model of the PV plant. We introduce here the selected PV model, which is the one-diode five-parameter model from \cite{desoto}, shown in Fig. \ref{Figure:PVcirc} for a single PV cell. Its main advantage is that it only requires datasheet information, normally available from the panel manufacturer. 
The adopted model $f(\cdot)$ describes the relationship between the DC voltage $v$, current $i$, solar irradiance $S$, and cell temperature $T$ for a PV panel composed of $ n_p $ cells in parallel and $ n_s $ in series. The model is:
\begin{align}
\begin{aligned}
&f(v,i,T,S) = 0 = I_{p}(T,S)n_p + - \frac{v+R_{s} i \frac{n_s}{n_p}}{R_{p}(S) \frac{n_s}{n_p}} - i+\\
& - i_D(T) n_p \left[\exp{\left(q\frac{v+R_{s} i \frac{n_s}{n_p}}{n_r k T n_s}\right)}-1\right]  
 , \label{eqref:diode}
\end{aligned}
\end{align}
where $k, q$ are physical constants and stand for diode Boltzmann constant and electron charge, respectively. The parameters $R_s, R_p, I_p, i_d, n_r $  respectively denote the series resistance, shunt resistance, light current, saturation current and diode ideality factor. Let the notation $*$ denote values at {Standard Test Condition} (STC), i.e. temperature $T^{*}=\SI{25}{\celsius}$ and irradiance $S^*=\SI{1000}{\watt\per\square\meter}$, the five parameters at different conditions \textit{(T,S)} are calculated as:
\begin{align}
& R_{s} = R^{*}_{s} \label{Rs} \\
& R_{p} = R^{*}_{p} \frac{S^{*}}{S} \label{eqref:Rsh} \\
& I_{p} = \left(I^{*}_{p}+\alpha\left(T-T^{*}\right)\right) \dfrac{S}{S^{*}} \label{Iph} \\
& n_r=n^{*}_r \label{n} \\
& i_{D} = i^{*}_{D}[T/T^{*}]^3\exp{\left(E^{*}_{g}/kT^{*}-E_{g}/kT\right)} \label{Id} \\
& E_g = 1.17-4.73 \times 10^{-4}\frac{T^2}{T+636}, \label{Eg}
\end{align}
where $E^{*}_{g}$ is the band gap energy (eV) at $T^*$, while $ R^{*}_{s}, R^{*}_{p}, I^{*}_{p},i^{*}_{D}, n^{*}_r$  are the parameters at STC, calculated with the procedure detailed in \cite{Laudani2013} by using the following datasheet information: the open circuit voltage $v^{*}_{OC}$, the short circuit current $i^{*}_{SC}$, the voltage and the current at the maximum power $v^{*}_{MP}$ and $i^{*}_{MP}$, the absolute temperature coefficients of the open circuit voltage $\beta$ and short circuit current $\alpha$. 

\section{Estimators of Solar Irradiance}\label{Estimators}
This section describes three selected methods to estimate the irradiance received by a PV array. The inputs are the DC voltage, DC current, and the cell temperature while the single diode model introduced in Section~\ref{PVmodel} is used to describe the PV system, for all the three cases. The output of the model (estimated irradiance, $ \hat{S} $) is then used to reconstruct the maximum power as described in Section~\ref{power}. We assume that the measured cell temperature is representative for the whole plant, in other words we assume a uniform temperature distribution. {\color{black}{To test the fairness of this modelling assumption, two identical temperature sensors were installed in different parts of the plant. They recorded an average temperature difference lower than $0.4 ^{\circ} $C and a maximum value of $ \approx $\SI{0.8}{\celsius}. Such a value determines a difference in the estimated maximum power of  $ \approx$0.5\%, thus making the assumption acceptable for the considered rooftop PV installation. In case of larger scale PV plants, this assumption should be re-evaluated, and the single-diode model could be replicated considering more temperature sensing points.
The sensitivity of the models to temperature measurement errors is further discussed in Section \ref{noiseMPPT}.}} 
		

\subsection{Analytical Formulation}
The irradiance is calculated analytically and in a closed-form by substituting equations~\eqref{Rs}-\eqref{Id} into \eqref{eqref:diode} and solving for $ S $. Formally, it is:
\begin{align}
\scalebox{0.78}{$
	\hat{S}_{A}=\dfrac{i+i^{*}_{D}n_p[T/T^{*}]^3\exp{\left(E^{*}_{g}/kT^{*}-E_{g}/kT\right)}\left[\exp{\left(q\frac{v+R^{*}_{S}in_s/n_p}{n_rkTn_s}\right)}-1\right]}{\frac{1}{S^{*}}\left[n_p\left(I^{*}_{p}+\alpha(T-T^{*})\right)-\frac{v+R^{*}_{S}in_s/n_p}{R^{*}_{p}n_s/n_p}\right]}$}
\end{align}
where $v$, $i$ and $T$ are measured quantities and $ \hat{S}_{A} $ is the inferred irradiance.

\subsection{Immersion And Invariance}\label{II}
Authors of \cite{carrasco2014} design an estimator exploiting the fact that the \textit{i-v} characteristics described by Eq. \eqref{eqref:diode} can be re-parametrized to show a monotonic behavior. The estimator is based on the principles of immersion and invariance, originally described in \cite{astolfi2007}.
The re-parameterization, based on the model in Eq. \eqref{eqref:diode}, is as follows. They  define  a \textit{measurable} signal $ y(t) $:
\begin{align}
&y(t)=i(t)- F(i(t), v(t), T(t))
\end{align}
where,
\begin{align}
&F(i, v, T)=I_0(T)\exp\left(\frac{C_1}{T}\left(v+C_2i\right)-1\right)\\
&I_0(T)=-C_6T^3\exp\left(C_7-\frac{C_8}{T}+\dfrac{C_9T}{T+C_{10}}\right).
\end{align}
$ C_i $ are constant values that can be calculated from single-diode equations and found in \cite{carrasco2014}.
The only difference is in the definition of constant $ C_3 $ since Authors of \cite{carrasco2014} 
consider a proportional relation between $ R_p $ and $ S $, while we here consider inverse proportionality, as expressed in Eq. \eqref{eqref:Rsh} and proposed in \cite{desoto}.

Then, they express the nonlinear regression form as:
\begin{align}
&y(t)=\Phi(S,t)\\
&\Phi(S,t)=S\left(C_4+C_5T(t)\right)- C_3/S\left(v(t)+C_2i(t)\right)
\end{align}
with $ \Phi(S,t) $ strictly monotonically increasing with $ S $.
At this point, the immersion and invariance estimator states:
\begin{align}
\dot{\hat{S_I}}=\gamma\left[y-\phi(\hat{S_I})\right]
\end{align}
and $ \gamma >0 $ ensures: 
\begin{align}
\lim_{t\to\infty}  \hat{S_I}(t)=S
\end{align}
where $ \hat{S_I} $ is the estimated irradiance.
Performance depends on the value of parameter $ \gamma $ that should be selected as a trade-off between convergence speed and noise filtering. 
\subsection{Extended Kalman Filter}
We propose to apply Kalman filter to estimate the irradiance as a function of voltage, current, and temperature measurements\footnote{As known from the existing bibliography, Kalman estimation consists in reconstructing the state of a system with noisy measurements by integrating the knowledge of the process which generated them.}. 
{\color{black}The advantage of a KF over a conventional low pass filter is that, by exploiting the knowledge of the process model, it achieves to filter out system disturbances and measurements noise on the whole spectrum of the state variables.}
The prerequisite to apply Kalman filtering is the knowledge of the system model and covariance matrices of system noise and measurements. To this end, we exploit the results from a previous work of the Authors \cite{Scolari2016116}, where it is shown that the irradiance evolution in the few seconds time scale can be captured with a persistence model plus a random variation from an identifiable pdf (probability density function), which is function of certain data features.

%
Let the state $x_k=S_k$ be the irradiance. At each discrete time $ k $, the measurements $ v_k, i_k$ and $ T_k$ are linked to the state by the nonlinear relationship $f(\cdot)$ in Eq. \eqref{eqref:diode}. Let  $f_1(\cdot), f_2(\cdot), f_3(\cdot)$ denote the function $f(\cdot)$ solved for voltage, current and temperature:
\begin{align}
&v_k=f_1(x_k, i_k,T_k) \label{eq:imp0}\\
&i_k=f_2(x_k,v_k, T_k)  \label{eq:imp1}\\
&T_k=f_3(x_k, v_k, i_k)\label{eq:imp2}.
\end{align}
The observation vector $\boldsymbol{y}_k=[v_k, i_k, T_k]^T$ is approximated as:
\begin{align}
\boldsymbol{y}_k \approx \boldsymbol{H_k} x_k + \boldsymbol{D_k}, \label{eq:biased_observer}
\end{align}
where $\boldsymbol{H}=[H_1, H_2, H_3]^T$ and $\boldsymbol{D}=[D_1, D_2, D_3]^T$ are from first order Taylor expansions of $f_1, f_2, f_3$. For example, for the case of $f_1$, they are:
\begin{align}
& v_k \approx f_1(\cdot) + f_{1,x}(\cdot) (x_k-a_k)\\
& H_1 = f_{1,x}(\cdot)\\
& D_1=f_1(\cdot) - f_{1,x}(\cdot)\cdot a_k
\end{align}
where $f_1(\cdot)$ and $f_{1,x}(\cdot)$ denotes the function and its first order derivative calculated in the point $a_k, i_k, T_k$, with $a_k$ is the irradiance value around which linearizing (assumed as the last available estimate, i.e. $a_k=x_{k-1}$) and $i_k$ and $T_k$ are both from measurements. 

The state-space formulation of the system model is:
\begin{align}
& x_k=F_{k-1}x_{k-1} + w_{k-1}, && w_{k-1} \in N(0,Q_k) \\
& \boldsymbol{y}_k = \boldsymbol{H_k} x_k + \boldsymbol{u}_k, && \boldsymbol{u}_k \in N(\boldsymbol{0},R_k) .\label{eq:KF1}
\end{align}
where  $F_{k-1}=1$ is the (scalar) system matrix, $Q_k$ is the system noise variance, and $R_k$ the $3 \times 3$ measurement noise covariance matrix.

The variance $Q$ is computed by applying the method described in \cite{Scolari2016116} and {\color{black} summarized for the sake of clarity in the Appendix.}

The covariance matrix  of measurements noise $R$ is a diagonal matrix $R=\text{diag}(\sigma^2_1, \sigma^2_2, \sigma^2_3)$. Measurements are assumed to be uncorrelated. The variance components are calculated assuming that the tolerance of the sensors (from datasheet values as specified in Section \ref{setup}) corresponds to the the 3-sigma level of a Gaussian distribution with zero mean, i.e. $\sigma_i = t_i/3, i=1,2,3$, where $t_i$ is the tolerance of the instrument $i$.

Once $\boldsymbol{H_k}$, $Q_k$ and $R_k$ are known from the procedures described above, the expected value $\hat{x} = \mathrm{E}[x_k]$ and variance $P_k = \mathrm{Var}[x_k]$  of the estimation are:
\begin{align}
&\hat{x}_k=(I-\boldsymbol{K_k}\boldsymbol{H_k})(F_{k-1}\hat{x}_{k-1})+ \boldsymbol{K_k}\boldsymbol{y}_k\\ \label{eq:est}
&	P_k=(I-\boldsymbol{K_k}\boldsymbol{H_k})(F_{k-1}P_{k-1}F_{k-1}^T+ Q_{k-1}),
\end{align}
where $I$ is the identity matrix, and $ \boldsymbol{K_k} $ is the Kalman gain:
\begin{align}
& \boldsymbol{K_k}=P_{k-1}\boldsymbol{H_k}^T(\boldsymbol{H_k}P_{k-1}\boldsymbol{H_k}^T +R_k)^{-1} \label{eq:KF3}.
\end{align}
We note that the linearization of the observer equation leads to an extended Kalman filter (EKF) formulation.

\section{Maximum Power Computation}\label{power}
Once the irradiance is estimated from any of the methods proposed in Section \ref{Estimators}, it is then used, together with $T$, to compute the maximum power output according to the following procedure:
\begin{itemize}
	\item as defined in \cite{IEC60891}, the open circuit voltage $v_{OC}$ is:
	\begin{align}
	v_{OC}=v^{*}_{OC}\left(1+\beta(T-T^{*})\right) + V_{t}n_rn_s\ln\left(\frac{\hat{S}}{S^{*}}\right)
	\end{align}
	where $V_t=kT/q$ is the thermal voltage;
	\item we determine the \textit{i-v} curve by applying Eq.~\eqref{eqref:diode} for values of the DC voltage  $v$ varying between 0 and $ v_{OC} $\footnote{The diode equation, (\ref{eqref:diode}), can be solved numerically or using the explicit formulation based on the Lambert function, described in \cite{jain2004exact}, that gives an exact analytical solution for $i$ as a function of $v$ and it is computationally more efficient.};
	\item  the maximum power of the PV module is computed as the maximum product $i \cdot v$ from the \textit{i-v} curve; 
	\item  the array maximum power is obtained by multiplying the module maximum power by the number of modules.
\end{itemize}

\section{Experimental Setup}\label{setup}
\begin{figure}[h!]
	\centering
	\includegraphics[scale=0.6]{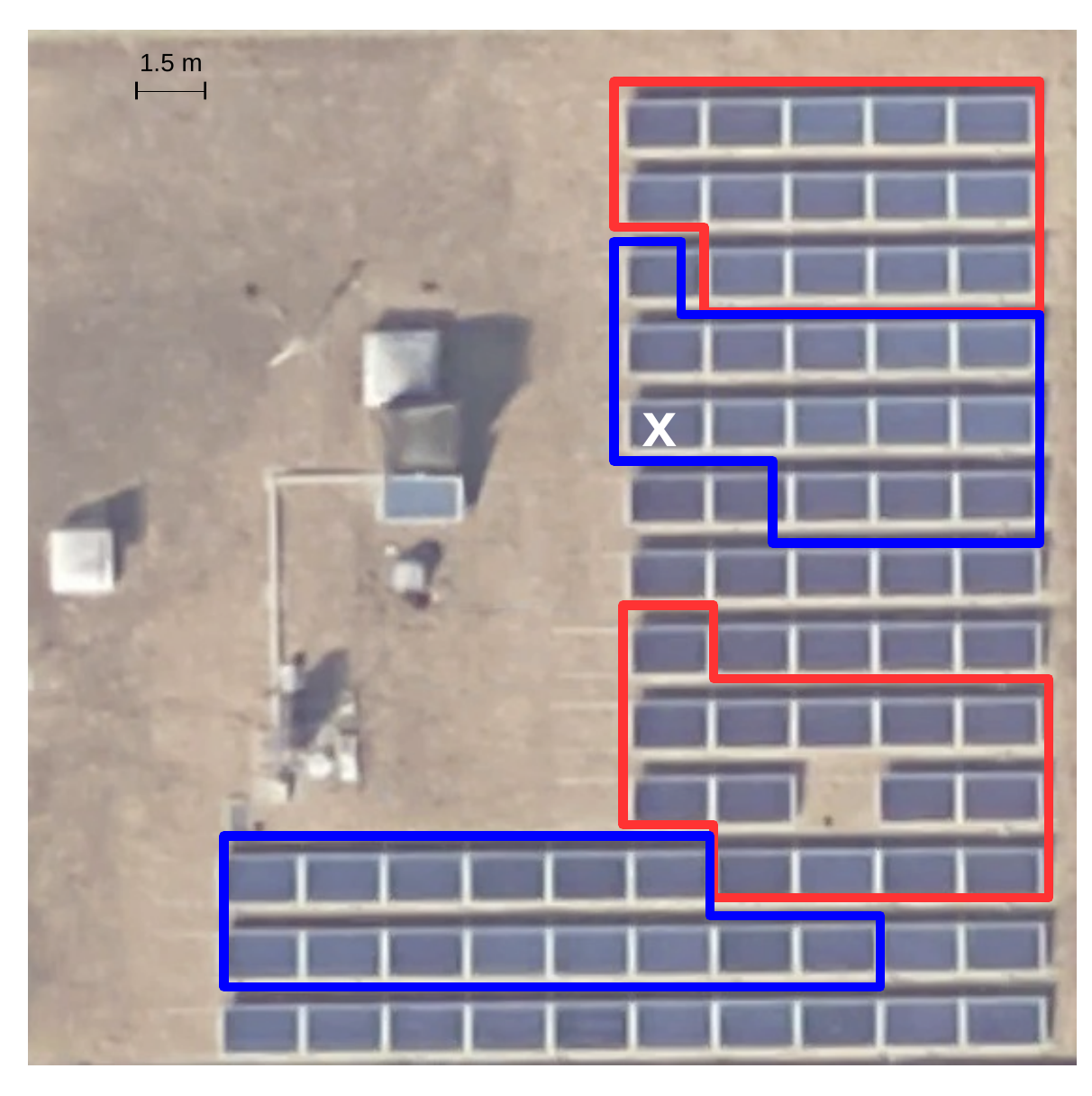}
	\caption{PV installation at the EPFL laboratory (GPS coordinates 46.52-N, 6.56-E). The strings indicated with blue color are connected to C1 and the strings indicated with red color are connected to C2. The white cross indicates the panel where the temperature sensor and the pyranometer are installed.}
	\label{Figure:pv}
\end{figure}

The considered experimental setup is a 14.3~kWp rooftop PV installation of the EPFL Romande Energie solar park in Switzerland and consists in two identical subsystems. {\color{black} Each subsystem is composed of one couple of strings (each with with 14 ECSOLAR 255~W Polycrystalline modules in series) connected to a three-phase DC/AC power converter. Panels are south facing and with 10 degrees tilt from the horizontal plane. The whole system is shown in Fig.~\ref{Figure:pv}, where the individual subsystems are marked with red and blue color.} The two converters, denoted by C1 and C2, are of the same commercial model and work independently. They can operate in MPPT mode {\color{black}{(using the Perturb and Observe method)}} or follow a specific active power external request (non-MPPT mode). This dedicated setup allows validating the capability of the proposed estimators also when the plant operates in non-MPPT mode, thanks to sending specific set-points requests to C1 while leaving C2 in MPPT mode (its output power is considered as the ground truth maximum power value).  {\color{black} The equivalent behavior of the two converters has been tested experimentally by running both of them in MPPT mode: in this case, their power output differ of less than 0.2\%. }
DC currents are measured with LEM LF 205-S current transducers with an accuracy of   $ \pm  $ 0.2\%. DC voltages are measured with LEM-CV 3-100 voltage transducers with an accuracy of $ \pm $0.5\%. 
Panel global normal irradiance (GNI) is measured with a silicon-cell device to limit the spectral mismatch due to the different spectral response between the modules and the sensor, \cite{dunn2012}. Thus, an Apogee SP-230 all-season pyranometer is installed in the same location of the plant. The pyranometer has an error  of $ \pm$2\% and $\pm$5\% at solar zenith angles of 45 and 75 degrees, respectively. 
The cell temperature is measured using a TSic303, a sensor with $\pm$\SI{0.5}{\degreeCelsius} accuracy, installed on the rear surface of the panel, as done in \cite{smith2011}. 
In order to correct the temperature readings accounting for the thermal resistance of the support material, we follow the procedure described in \cite{king2004photovoltaic} and add a positive offset with magnitude n$\cdot$\SI{3}{\degreeCelsius}, where $ n $ is dimensionless irradiance (we here use $ \SI{1000}{\watt\per\square\meter} $ as base quantity). The value of $ n $ is calculated using the irradiance estimated at the previous time-step,  which is a fair assumption since our estimations are at high time resolution  and the temperature dynamics are slower than the irradiance ones.
 {\color{black}{The location of the pyranometer and temperature sensor is marked with a white cross in Fig.~\ref{Figure:pv}. }}

{\color{black}{Analog measurements of voltage and current are acquired at 20~kHz  with an 18 bit analog-to-digital converter (ADC, NI CompactRIO 9068 equipped with a 9215 module), while irradiance and temperature are sampled 20~kHz with a NI sbRIO 9625 with a 16 bit ADC. All the measurements are then sampled at 1~s and saved in a time series database. The two acquisition devices mentioned above are synchronized, and the two groups of measurements are with a time jitter of 0.5~s at most. Only daylight data are selected for the analysis, i.e. solar elevation larger than 3 degrees. }}

\section{Results}\label{results}
In this section, the performance of the methods in Section~\ref{Estimators} is compared in terms of quality of the estimations of the maximum power they provide. Summarizing from the previous sections, this consists in i) computing the irradiance using the analytical (Analytical), immersion and invariance (I$\&$I), and extended Kalman Filter (EKF) methods and ii) estimating the maximum  power with the procedure described in Section \ref{power}. The quality of the maximum power estimations is assessed by comparing them against the measured power flow (ground truth value) of converter C2 that is always in MPPT mode. As a further benchmark, we include the maximum power estimation performed by using measurements from the pyranometer: this consists in feeding the procedure for the maximum power estimation in Section \ref{power}  with GNI readings and cell temperature measurements. 
The section is organized as follows.
 Section~\ref{metrics} presents the metrics used for the comparison. Sections~\ref{MPPT_cur} and \ref{MPPT_res} compare the performance of the proposed methods for maximum power estimation for non-MPPT and MPPT conditions, respectively. Finally, in Section~\ref{noiseMPPT} the robustness of the proposed estimators in terms of rejection against measurements noise is assessed by decreasing their signal-to-noise ratio. Two days are considered for the analysis: a clear-sky  (9 of September 2016) and a partly cloudy day (12 of September 2016). 

\subsection{Metrics}\label{metrics}
Let ${P}_t$ be the ground truth maximum power value at the time interval $t=1,\dots,T$, where $T$ is the number of samples, $\widehat{P}_t$ the estimation, $\overline{P}$ the average over time of the ground truth values in the interval $t=1,\dots,T$. Three metrics are used to characterize the performance of the proposed techniques, the normalized root mean squared error (nRMSE), the normalized maximum error ($ \mathrm{\text{Err}_{\text{max}}} $) and the  normalized mean error (nME):
\begin{align}
&\text{nRMSE}=\dfrac{1}{\overline{P}}\sqrt{{\sum\limits_{t=1}^m (\hat{P_t}-P_t){\color{black}^2}}/{m}}, \\
& \text{Err}_{\text{max}}  = \text{max}\left\{{\left|\hat{P_t}-P_t \right|}/{\overline{P}},\ t=1,\dots,m\right\}, \\
& \text{nME}=\dfrac{1}{m}\sum\limits_{t=1}^m {(\hat{P_t}-P_t)}/{\overline{P}}.
\end{align}

		\subsection{Maximum Power Estimation in non-MPPT Conditions}\label{MPPT_cur}
\begin{figure*}[ht!]
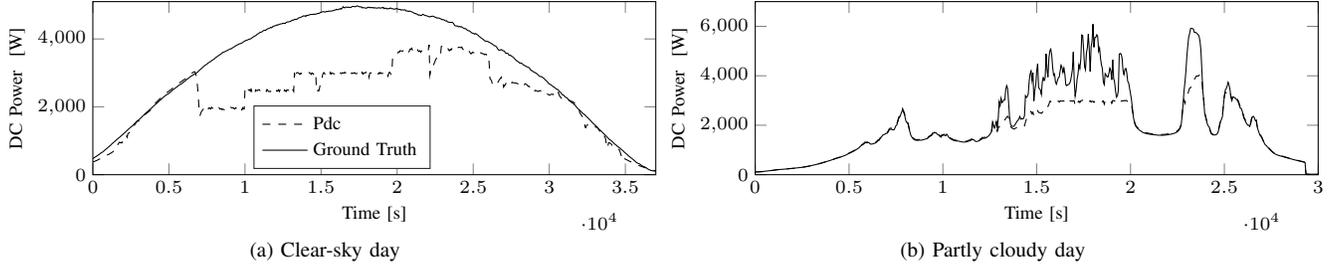

				\subfloat[Clear-sky day]{
					\small
					\footnotesize
					\scriptsize
					\label{fig:Real_curtailedcs}
					\input{fig/Real_curtailed_b.tex}} 
				\subfloat[Partly cloudy day]{
					\small
					\footnotesize
					\scriptsize
					\label{fig:Real_curtailedpc}
					\input{fig/Real_curtailed.tex}} 
				\caption{Maximum power (Ground Truth) and curtailed power (Pdc) are shown. Measurements come from two identical converters (same technology) that are working under equivalent conditions but different modes (i.e. MPPT for C2 and non-MPPT/curtailed for C1).} 
				\label{fig:Real_curtailed}
			\end{figure*}

\begin{figure*}[ht!]
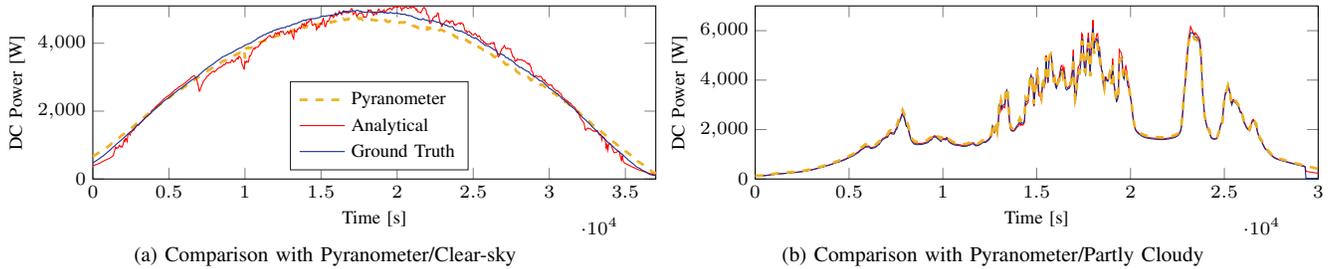

							\subfloat[Comparison with Pyranometer/Clear-sky]{
							\small
							\footnotesize
							\scriptsize
							\label{fig:Compacs}
							\input{fig/clear-sk_pyr.tex}} 
							\subfloat[Comparison with Pyranometer/Partly Cloudy]{
								\small
								\footnotesize
								\scriptsize
								\label{fig:Compapc}
								\input{fig/Companew.tex}} 
						\caption{Comparison between the ground truth maximum power, the estimation using the analytical formulation and the one  starting from the irradiance sensed by a pyranometer.} 
						\label{fig:Compat}
					\end{figure*}

		Tables \ref{Tab:Curtailederrorcs} and \ref{Tab:Curtailederror} show the performance of the proposed techniques and of the pyranometer-based estimations, when the power output of converter C1 is curtailed as shown in Fig. \ref{fig:Real_curtailed}, for a clear-sky and partly cloudy day respectively. As explained in Section~\ref{II}, performance of the I$\&$I estimator depends on the value of parameter $ \gamma $ that should be selected as a trade-off between good convergence and noise filtering. Authors of \cite{carrasco2014} suggest values in the range 0.1-1, thus we first fix $ \gamma $ equal to 0.7.
					\begin{figure*}[ht!]
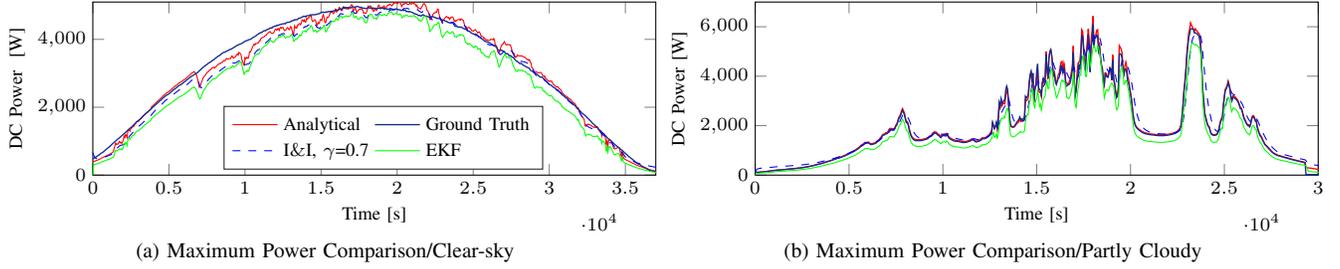

				\subfloat[Maximum Power Comparison/Clear-sky]{
					\small
					\footnotesize
					\scriptsize
					\label{fig:allEst}
					\input{fig/allEst.tex}} 
				\subfloat[Maximum Power Comparison/Partly Cloudy]{
					\small
					\footnotesize
					\scriptsize
					\label{fig:allEstpc}
					\input{fig/allEstpc.tex}} 
				\caption{Comparison between the ground truth maximum power  and the reconstructed maximum power using the analytical, the EKF, and the I$\&$I estimators.}
				\label{fig:allEstt}
			\end{figure*}
		Performance denotes that the methods are able to reconstruct the maximum output power 
		starting from any operating point, as shown in Fig. \ref{fig:allEstt}. Indeed, it is possible to reconstruct the  maximum power even when starting from curtailed conditions. 
		The analytical method outperforms the other estimators, during both clear-sky and partly cloudy conditions. {\color{black}{ For the pyranometer-based estimations,  larger errors are likely generated when GNI measurements are not representative of the conditions of the whole plant.
	 Figures \ref{fig:Compacs}  and \ref{fig:Compapc} compare the ground truth maximum power values against the analytical and pyranometer-based estimations for the clear-sky and partly cloudy day, respectively. For a more detailed comparison we refer to numerical results in Tables \ref{Tab:Curtailederrorcs} and \ref{Tab:Curtailederror}.}}
	
\begin{figure*}[ht!]
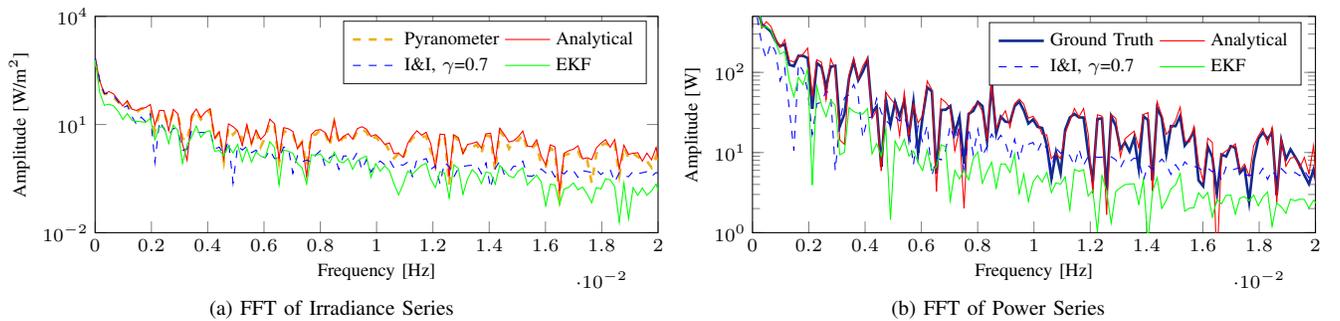

						\subfloat[FFT of Irradiance Series]{
							\small
							\footnotesize
							\scriptsize
							\label{fig:fft_irr}
							\input{fig/fft_irr.tex}} 
						\subfloat[FFT of Power Series]{
							\small
							\footnotesize
							\scriptsize
							\label{fig:fft_p}
							\input{fig/fftb.tex}} 
						\caption{The FFT single-sided amplitude spectrum of the estimated and measured irradiances is shown as a function of the frequency in a semilogarithmic scale in Fig. \ref{fig:fft_irr}. The same spectrum is shown for the estimated and ground truth maximum power values in Fig. \ref{fig:fft_p}.}
						\label{fig:fft}
					\end{figure*}

		Tables \ref{Tab:Curtailederrorcs} and \ref{Tab:Curtailederror} also show that the EKF and the I$\& $I (with $ \gamma $=0.7) estimators have worse performance than the others, especially in the partly cloudy case. This fact is justified by their smaller bandwidth. In support to this, we compare in Figures~\ref{fig:fft_irr} and \ref{fig:fft_p} the amplitudes of the fast Fourier transform (FFT) of the ground truth values and estimations, for the irradiance and maximum power respectively. We consider 6200 samples taken from a period with large power variations, in the central part of the cloudy day.
		 As visible from Fig.~\ref{fig:fft_irr}, only the analytical estimation is able to track the irradiance fluctuations registered by the pyranometer, especially at high frequency. Similarly, in Fig.~\ref{fig:fft_p}  we observe that the EKF and I$\&$I-based power estimations tend to filter out fast but significative dynamics, thus determining poorer performance. On the contrary, the analytical formulation is able to follow the ground truth series even at high frequencies.
		A further analysis we propose, consists in  gradually increasing the value of $ \gamma $ in order to favor the I$\&$I performance over the noise filtering. Results show that values of $ \gamma $ in the range 20-200 (for the partly cloudy day) and 5-200 (for the clear-sky day) allow achieving performance very close to the analytical case, thus denoting that the choice of $ \gamma $ is critical and its selection may represent a drawback with respect to parameter-less methods as the analytical one. 
	 {\color{black}{The worse results of the EKF are due to the non-stationary of the irradiance time series, even when differentiated, which makes difficult to identify an exact model of the process.}}
		
		\begin{table}[!ht]
			\renewcommand{\arraystretch}{1.1}
			\centering
			\caption{Performance for a clear-sky day/non-MPPT.}
			\label{Tab:Curtailederrorcs}
			\scalebox{1.2}{
				\begin{tabular}{|c|c|c|c|}
					\hline
					&nRMSE [\%] &  $\text{Err}_{\text{max}} $ [\%] & nME [\%]\\
					\hline
					Analytical & 5.40  & 20.4 & -0.8\\
					\hline
					EKF & 11.53 & 21.2 & -10.4 \\
					\hline
					I$\&$I, $ \gamma$=0.7 & 7.46 & 20.1 & -4.57\\
					\hline
					Pyranometer & 6.65 & 15.1  & -2.76 \\
					\hline
				\end{tabular}}
			\end{table}
			
			\begin{table}[!ht]
				\renewcommand{\arraystretch}{1.1}
				\centering
				\caption{Performance for a partly cloudy day/ non-MPPT.}
				\label{Tab:Curtailederror}
				\scalebox{1.2}{
					\begin{tabular}{|c|c|c|c|}
						\hline
						&nRMSE [\%] &  $\text{Err}_{\text{max}} $ [\%] & nME [\%]\\
						\hline
						Analytical & 5.6  & 7.15  & 2.53 \\
						\hline
						EKF & 19.9 &30.0  &  -13.9\\
						\hline
						I$\&$I, $ \gamma$=0.7 & 19.9 & 23.0 & 5.17\\
						\hline
						Pyranometer & 9.16 & 32.8 & 7.14\\
						\hline
					\end{tabular}}
				\end{table}
					
				\subsection{Maximum Power Estimation in MPPT Conditions}\label{MPPT_res}
				Tables \ref{Tab:MPPTerrorcs} and \ref{Tab:MPPTerror} show the performance of the proposed techniques and of the pyranometer-based estimations when the system is in MPPT mode for the whole period, for a clear-sky and a partly cloudy day, respectively. The analysis confirms what observed for the non-MPPT case. Results from the pyranometer estimations are independent of the PV system state and are therefore identical for the MPPT and non-MPPT analysis.  On the other hand, the analytical method is the most sensitive to the operating point conditions and its performance improves when the system is close to the maximum power point.
				As far as the I$\&$I method is concerned, same results of Section~\ref{MPPT_cur} apply regarding the tuning of parameter $ \gamma $.

				\begin{table}[!ht]
					\renewcommand{\arraystretch}{1.1}
					\centering
					\caption{Performance for a clear-sky day/MPPT.}
					\label{Tab:MPPTerrorcs}
					\scalebox{1.2}{
						\begin{tabular}{|c|c|c|c|}
							\hline
							&nRMSE [\%] &  $\text{Err}_{\text{max}} $ [\%] & nME [\%]\\
							\hline
							Analytical & 0.51  & 1.5 & 0.37\\
							\hline
							EKF & 11.53 & 18 & -13.3 \\
							\hline
							I$\&$I, $ \gamma$=0.7 & 4.21 & 6.8 & -3.51 \\
							\hline
							Pyranometer & 6.65 & 15.1 &-2.76 \\
							\hline
						\end{tabular}}
					\end{table}
					
					\begin{table}[!ht]
						\renewcommand{\arraystretch}{1.1}
						\centering
						\caption{Performance for a partly cloudy day/MPPT.}
						\label{Tab:MPPTerror}
						\scalebox{1.2}{
							\begin{tabular}{|c|c|c|c|}
								\hline
								&nRMSE [\%] &  $\text{Err}_{\text{max}} $ [\%] & nME [\%]\\
								\hline
								Analytical & 3.7  & 5 & 1.51\\
								\hline
								EKF & 19.9 & 35.1 &-14.1 \\
								\hline
								I$\&$I, $ \gamma$=0.7 & 19.9 & 22 & 4.06\\
								\hline
								Pyranometer & 9.16 & 32.8 & 7.14\\
								\hline
							\end{tabular}}
						\end{table}
					
				\subsection{Robustness Against Measurements Noise} \label{noiseMPPT}
				The EKF and I$\&$I-based estimations are expected to have improved rejection against measurements noise, thanks to integrating into the estimation process measurements which are progressively becoming available with time. In this section, we deliberately decrease the signal-to-noise ratio of the measurements with the objective of showing the break-even between the performance of the estimators. In other words, we want to investigate on the level of measurement noise after which the use of filter-based estimators is advisable compared to the analytical formulation.
				Original voltage, current and temperature measurements (which are already characterized by the original noise inherent the respective sensing devices, see Section \ref{setup}) are corrupted with an additive independent and identically distributed (i.i.d.) Gaussian noise with increasing values of standard deviation (STD). This analysis is carried out for each measurement in a separate fashion, namely noise is added to a single measurement (voltage, current or temperature) while keeping the others to their original level of noise.\footnote{In the EKF case, the element of matrix $ R $ corresponding to the noisy measurement is augmented with the value of the variance of the fictitious additional noise. Assuming that the variables are uncorrelated, we can say that the variance of the sum equals the sum of the variances.} 
				The original levels of noise and those artificial added are recapped in Table~\ref{Tab:STD}.
				
					\begin{table}[!ht]
						\renewcommand{\arraystretch}{1.1}
						\centering
						\caption{Standard Deviations (STD) of the Input Measurements}
						\label{Tab:STD}
						\scalebox{1.2}{
							\begin{tabular}{|c|c|c|}
								\hline
							 & Original noise (STD) &Added Noise (STD)\\
								\hline
								$ i $ & 0.55 & 0.3,0.45,0.55,0.7  \\
								\hline
								$ v$ & 0.23 & 0.3,0.71,1\\
								\hline
								$ T $ & 0.4 & 1.4,2.23,3.16\\
								\hline
							\end{tabular}}
						\end{table}

\begin{figure}[ht!]
					\subfloat[MPPT]{
						\small
						\footnotesize
						\scriptsize
						\label{fig:noise_current_mppt}
%
%
\begin{tikzpicture}

\begin{axis}[%
width=0.6\matlabfigurewidth,
height=0.52\matlabfigurewidth,
at={(0\matlabfigurewidth,0\matlabfigurewidth)},
scale only axis,
xmin=0.57,
xmax=0.89,
xlabel={Standard Deviations [A]},
ymin=5,
ymax=40,
ylabel={rRMSE [\%]},
axis background/.style={fill=white},
legend style={font=\tiny,at={(0.96,0.99)}, legend columns=2 legend cell align=left,align=left,draw=white!15!black}
]
\addplot [color=red,solid, width=1.0pt]
  table[row sep=crcr]{%
0.57	6.38\\
0.63	16.3\\ 
0.70	21.1\\ 
0.77	27.1\\  
0.89	32.0\\  
};
\addlegendentry{Analytical};

\addplot [color=blue,dashed, width=1.0pt]
  table[row sep=crcr]{%
0.57	20.13\\
0.63	20.1\\
0.70	20.2\\
0.77	20.4\\
0.89	20.34\\
};
\addlegendentry{I$\&$I,$ \gamma $=1};

\addplot [color=blue,densely dotted,width=1.0pt]
table[row sep=crcr]{%
	0.57	6.4\\
	0.63	8.9\\
	0.70	10\\
	0.77	11\\
	0.89    12\\
};
\addlegendentry{I$\&$I,$ \gamma_{opt} $};

\addplot [color=green,solid, width=1.0pt]
  table[row sep=crcr]{%
0.57	20.6\\
0.63	20.6\\
0.70	21.1\\
0.77	22\\
0.89	23\\
};
\addlegendentry{EKF};

\end{axis}
\end{tikzpicture}%
} 
					\subfloat[Curtailed]{
						\small
						\footnotesize
						\scriptsize
						\label{fig:noise_current_curtailed}
%
%
\begin{tikzpicture}

\begin{axis}[%
width=0.56\matlabfigurewidth,
height=0.52\matlabfigurewidth,
at={(0\matlabfigurewidth,0\matlabfigurewidth)},
scale only axis,
xmin=0.57,
xmax=0.89,
xlabel={Standard Deviations [A]},
ymin=5,
ymax=40,
ylabel={rRMSE [\%]},
axis background/.style={fill=white},
legend style={at={(0.4,1)}, legend cell align=left,align=left,draw=white!15!black}
]
\addplot [color=red,solid, width=1.0pt]
  table[row sep=crcr]{%
0.57 7.17\\
0.63 21.1\\
0.7 25.9\\
0.77 33.8\\
0.89 40.1\\
};

\addplot [color=blue,dashed, width=1.0pt]
  table[row sep=crcr]{%
0.57	    19.9\\
0.63	    19.9\\
0.7	    20.26\\
0.77	21.67\\
0.89	   21.0\\
};
\addplot [color=blue,densely dotted, width=1.0pt]
table[row sep=crcr]{%
	0.57	    7.2\\
	0.63	    11.5\\
	0.7	    12.3\\
	0.77	13.5\\
	0.89	   14\\
};

\addplot [color=green,solid, width=1.0pt]
  table[row sep=crcr]{%
0.57	19.9\\
0.63	21\\
0.7     21.3 
0.77	21.67\\
0.89	   21.0\\
};

\end{axis}
\end{tikzpicture}%
} 
					\caption{Noise is increased on the DC current.} 
					\label{fi
						g:currents}
				\end{figure}
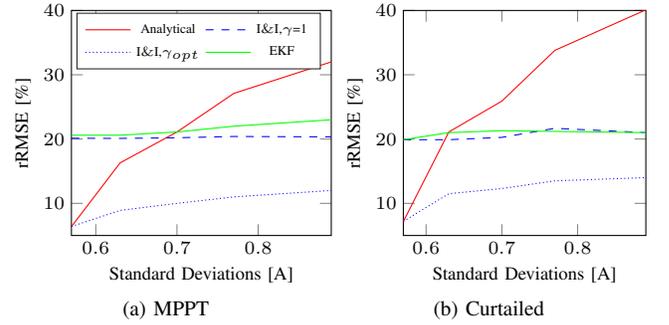

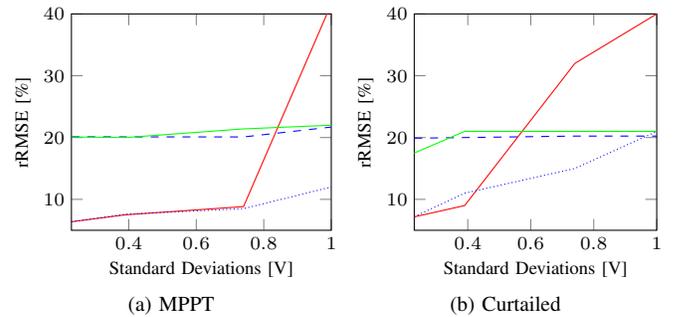
\begin{figure}[ht!]
					\subfloat[MPPT]{
						\small
						\footnotesize
						\scriptsize
						\label{fig:noise_voltage_mppt}
%
%
\begin{tikzpicture}

\begin{axis}[%
width=0.6\matlabfigurewidth,
height=0.5\matlabfigurewidth,
at={(0\matlabfigurewidth,0\matlabfigurewidth)},
scale only axis,
xmin=0.23,
xmax=1,
xlabel={Standard Deviations [V]},
ymin=5,
ymax=40,
ylabel={rRMSE [\%]},
axis background/.style={fill=white},
legend style={at={(0.6,0.95)}, legend cell align=left,align=left,draw=white!15!black}
]
\addplot [color=red,solid, width=1.0pt]
  table[row sep=crcr]{%
0.23	6.38\\
0.39	7.5\\
0.74	8.83\\
1.0	41.6\\
};

\addplot [color=blue,dashed, width=1.0pt]
  table[row sep=crcr]{%
0.23	20.13\\
0.39	20.11\\
0.74	20.1\\
1	21.7\\
};

\addplot [color=blue,densely dotted,width=1.0pt]
table[row sep=crcr]{%
	0.23	6.38\\
	0.39	7.6\\
	0.74	8.5\\
	1	12\\
};

\addplot [color=green,solid,width=1.0pt]
  table[row sep=crcr]{%
0.23	20.1\\
0.39	20.0\\
0.74	21.4\\
1.0	22\\
};

\end{axis}
\end{tikzpicture}%
} 
					\subfloat[Curtailed]{
						\small
						\footnotesize
						\scriptsize
						\label{fig:noise_voltage_curtailed}
%
%
\begin{tikzpicture}

\begin{axis}[%
width=0.56\matlabfigurewidth,
height=0.5\matlabfigurewidth,
at={(0\matlabfigurewidth,0\matlabfigurewidth)},
scale only axis,
xmin=0.23,
xmax=1,
xlabel={Standard Deviations [V]},
ymin=5,
ymax=40,
ylabel={rRMSE [\%]},
axis background/.style={fill=white},
legend style={at={(0.4,0.99)}, legend cell align=left,align=left,draw=white!15!black}
]
\addplot [color=red,solid, width=1.0pt]
  table[row sep=crcr]{%
0.23	7.17\\
0.39	9\\
0.74	32\\
1.0	40\\
};

\addplot [color=blue,dashed, width=1.0pt]
  table[row sep=crcr]{%
0.23	19.9\\
0.39 	20\\
0.74	20.24\\
1.0	20.24\\
};

\addplot [color=blue,densely dotted, width=1.0pt]
table[row sep=crcr]{%
	0.23	7.2\\
	0.39 	11\\
	0.74	15\\
	1.0	21\\
};

\addplot [color=green,solid, width=1.0pt]
  table[row sep=crcr]{%
0.23	17.5\\
0.39	21\\
0.74	21\\
1.0	21\\
};

\end{axis}
\end{tikzpicture}%
} 
					\caption{Noise is increased on the DC voltage.} 
					\label{fig:voltges}
				\end{figure}

\begin{figure}[ht!]
					\subfloat[MPPT]{
						\small
						\footnotesize
						\scriptsize
						\label{fig:noise_tem_mppt}
%
%
\begin{tikzpicture}

\begin{axis}[%
width=0.6\matlabfigurewidth,
height=0.5\matlabfigurewidth,
at={(0\matlabfigurewidth,0\matlabfigurewidth)},
scale only axis,
xmin=1,
xmax=3.3,
xlabel={Standard Deviations [K]},
ymin=5,
ymax=25,
ylabel={rRMSE [\%]},
axis background/.style={fill=white},
legend style={at={(0.6,0.96)}, legend cell align=left,align=left,draw=white!15!black}
]
\addplot [color=red,solid, width=1.0pt]
  table[row sep=crcr]{%
1	6.38\\
1.73	7.5\\
2.45	8.83\\
3.3	14.6\\
};

\addplot [color=blue,dashed, width=1.0pt]
  table[row sep=crcr]{%
1	20.1\\
1.73	20.0\\
2.45	21.4\\
3.3	22\\
};
\addplot [color=blue,densely dotted, width=1.0pt]
table[row sep=crcr]{%
	1	6.4\\
	1.73	7\\
	2.45	7\\
	3.3	 9\\
};

\addplot [color=green,solid, width=1.0pt]
  table[row sep=crcr]{%
1	19.9\\
1.73	21.0\\
2.45	22\\
3.3	22.3\\
};

\end{axis}
\end{tikzpicture}%
} 
					\subfloat[Curtailed]{
						\small
						\footnotesize
						\scriptsize
						\label{fig:noise_tem_curtailed}
%
%
\begin{tikzpicture}

\begin{axis}[%
width=0.56\matlabfigurewidth,
height=0.5\matlabfigurewidth,
at={(0\matlabfigurewidth,0\matlabfigurewidth)},
scale only axis,
xmin=1,
xmax=3.3,
xlabel={Standard Deviations [K]},
ymin=5,
ymax=25,
ylabel={rRMSE [\%]},
axis background/.style={fill=white},
legend style={at={(0.95,0.95)}, legend cell align=left,align=left,draw=white!15!black}
]
\addplot [color=red,solid, width=1.0pt]
  table[row sep=crcr]{%
1	5.6\\
1.73	8\\
2.45	8.5\\
3.3	19.9\\
};

\addplot [color=blue,dashed, width=1.0pt]
  table[row sep=crcr]{%
1	19.9\\
1.73	19.9\\
2.45	20.1\\
3.3	20.1\\
};
\addplot [color=blue,densely dotted, width=1.0pt]
table[row sep=crcr]{%
	1	5.6\\
	1.73	11.1\\
	2.45	13\\
	3.3	14\\
};

\addplot [color=green,solid, width=1.0pt]
  table[row sep=crcr]{%
1	19.9\\
1.73	21\\
2.45	21.2\\
3.3	22\\
};

\end{axis}
\end{tikzpicture}%
} 
					\caption{Noise is increased on the measured cell temperature.} 
					\label{fig:noiseT}
				\end{figure}
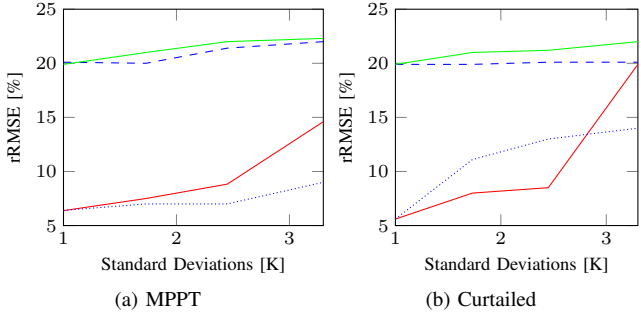

				The analysis considering the additional noise on voltage, current and temperature are shown in Figures~\ref{fig:noise_current_mppt}, \ref{fig:noise_voltage_mppt} and \ref{fig:noise_tem_mppt} for the MPPT case, and in Figures~\ref{fig:noise_current_curtailed}, \ref{fig:noise_voltage_curtailed} and \ref{fig:noise_tem_curtailed} for the non-MPPT case. The x-axis shows the total level of noise standard deviation (i.e. original noise level plus additive i.i.d. noise) while the y-axis the nRMSE associated with the maximum power estimation. The partly cloudy day is considered for the analysis. 
				For the I$\&$I we consider two cases: a first with $ \gamma $ fixed to 1 and a second case where $ \gamma $
				is tuned to optimize the performance ($ \gamma_{opt} $).
				Figures \ref{fig:currents},\ref{fig:voltges} \ref{fig:noiseT} show that the tuning of parameter $ \gamma $ allow the I$\&$I estimator to have  stable performance even for high level of noise. When the noise augments, performance of the analytical method deteriorates while the I$\&$I estimator with $ \gamma_{opt} $ maintains a low nRMSE.
				For the other cases (EKF and I$\&$I with $ \gamma=1 $)
				we can see that break-evens with respect to the analytical performance happen for an STD of $ \approx $0.68 for the current and $ \approx $0.82 for the voltage on the MPPT case. For the non-MPPT case, the break-even occurs for  an STD of $ \approx $0.63 for the current  $ \approx $0.58 for  the voltage. 
				 This break-even has, in general, a lower value in the curtailed case, for which the analytical estimation is more sensitive to the presence of measurement noise.
				Fig. \ref{fig:noiseT} shows that the analytical estimation is less sensitive to temperature noise, especially when working close to MPPT conditions.
				 The break-even is reached for very high STD, corresponding to an STD  of $ \approx $3.3 for the non-MPPT case.

\section{Improvement of data-driven forecasting methods}\label{application}
{\color{black}{We show how the discussed estimators are applied to treat historical production time series of a PV plant to filter out exogenous control action components (e.g. due to curtailment) from a training data set.  The objective is to show how the proposed estimators can be used to improve the performance of a machine learning-based forecasting method for maximum power prediction. For this purpose, we select a scenario where the PV power is curtailed, thus historical power measurements are not representative of the maximum available power from the plant.
As a forecasting tool, we implement an artificial neural network (ANN), a method often advocated in the existing literature in application to data-driven point predictions \cite{review2016}. We consider its prediction performance in two cases. First, when the ANN is trained by using raw production measurements (direct forecast, DF); this means that the training can contain values that are not representative of the maximum available power. Second, the training time series is generated with the proposed estimators and thus the training data always approximate the maximum available power  (filtered forecast, FF).}} This latter case implies the implementation of the analytical, EKF, $ I$\&$I $ estimators, and the pyranometer-based one to reconstruct the maximum power. For the $ I$\&$I $ we fix $ \gamma $=10. 
The fitting is performed considering 10 neurons and one hidden layer and using as training algorithm the Matlab implementation of the Levenberg-Marquard back-propagation function, \cite{LM}. The considered prediction horizon is 5 minutes.
The dataset consists of 35 days of experimental data (generated with the setup described in Section~\ref{setup}), 30 of which are used for the training phase and 5 for the testing. The training data set contains 12 days where the PV power output is curtailed according to a random pattern by controlling the active power set-point of the experimental power converter. The same applies for 2 days of the testing data set. For the remaining period, the power converter is left in MPPT mode.
The metric used to evaluated point predictions performance is the normalized mean absolute error (nMAE):
\begin{align}
\color{black}nMAE = \dfrac{1}{M}\sum\limits_{t=1}^M {(\tilde{P}_{t+1|t}-P_{t+1})}/{\overline{P}}
\end{align}
where  $\tilde{P}$ is the one-step-ahead prediction, $P$ is the realization, $M$ is the number of measurements in the testing dataset and $ \overline{P} $ their average value.
We remind here that the aim of this comparison is not to assess the skills of the forecasting method, rather showing the advantage introduced by the proposed pre-filtering approaches in improving time-series based forecasting tools. 
Results are summarized in Table~\ref{Tab:ANN-MAE}. They show that reconstructing the maximum power with the proposed techniques is always beneficial  when curtailment strategies are adopted: the DF, that simply uses past row measurements of the output power, produces the largest nMAE. Performance of the other methods is in-line with what obtained in the estimation comparison in Section \ref{results}.

\begin{table}[!ht]
	\renewcommand{\arraystretch}{1.1}
	\centering
	\caption{Normalized Mean Absolute Error }
	\label{Tab:ANN-MAE}
	\scalebox{1.2}{
		\begin{tabular}{|c|c|}
			\hline
			&{\color{black}{ nMAE [\%]}}\\
			\hline
			DF   & 19.7 \\
			\hline
			FF, Analytical & 9.0\\
			\hline
			FF, EKF & 14.7\\
			\hline
			FF, I$\&$I $ \gamma=10 $ &  9.3 \\
			\hline
			FF, Pyranometer &  11.4  \\
			\hline
		\end{tabular}}
	\end{table}
	
	%
	%
	%
	%
	%
	%

\section{Conclusion}\label{conclusion}
We have analyzed three model-based methods to estimate the irradiance received by a PV system  from  measurements of the system DC current, voltage and cell temperature. The estimators are  applied to reconstruct the maximum DC power output of a PV plant independently of the fact it operates in MPPT mode or under curtailment regimes. {\color{black}{Two of them, based on an analytical closed-form solution and on the immersion and invariance principles are taken from the literature while the third, based on the extended Kalman Filter, is originally proposed by the Authors.}} The estimation performance is evaluated by using measurements from an experimental setup and benchmarked against pyranometer estimations. Results show that:
\begin{itemize}
	\item the considered estimators can reconstruct successfully the theoretical maximum power output of PV installations even when the plants operate in non-MPPT mode;
	\item when estimating the peak power of PV systems,  the considered approach can outperform pyranometer-based estimations;
	 	\item {\color{black}{for noise levels compatible with commercial sensors, the analytical estimator showed similar or better performance and bandwidth than the immersion and invariance and Kalman filter-based estimators,  with the advantage of being parameter-less. If the available measurements are characterized by a high level of noise (STD higher than  $ \approx $0.6 for voltage and current and than $ \approx $3 for the temperature), the use of filter-based strategies is advisable since they are able to delete noisy observations while accounting for the structure of the process. In particular, the I\&I method is able to maintain low nRMSE but a preliminary tuning of $ \gamma $ is required.}}

\end{itemize}

{\color{black} { The proposed methods can be implemented in parallel to a machine learning maximum power forecast tool to aid its short-term prediction capabilities. This is possible with a simple setup that includes a temperature sensor, and voltage and current measurements that are generally available from the converter monitoring system. 

We expect that the results here presented can be of interest in the context of MPPT developing and monitoring applications. For example, PV array models are used to continuously monitor inverter's MPPT characteristics, in particular during high irradiance variations, \cite{king2007}. Further investigation in this direction can be part of a future work.
		
Future research will also consider the effects of fault occurrence and degradation processes.  For example, adaptive model identification with periodical re-training of model parameters can be implemented to account for degradation. Furthermore,  peer-to-peer collaborative strategies, exploiting estimations from neighboring PV installations, can be used to identify lack of homogeneity in production patterns and faults.}}  
\appendix
 {\color{black}{The following describes how to compute the variance \textit{Q} for the Estimated Kalman Filter-based irradiance estimation. In summary, it consists in grouping $ N $ historical values of the differentiated irradiance time series ($  \Delta{S} $) into clusters according to the value of selected data features:
 		\begin{itemize}
 			\item the average irradiance value on a mobile window of length $ n $ considering the most recent data points:
 			\begin{align}
 			M_{i}=\frac{1}{n}\sum\limits_{j=i-n}^i \Delta{S_{j}},\;\; i=n+1, \dots, N \label{average}
 			\end{align}
 			
 			\item the irradiance variability:
 			\begin{align}
 			V_{i}=\sqrt{\frac{1}{n}\sum\limits_{j=i-n}^i { (\Delta{S_{j}}- \Delta{S_{j-1}})}^2}, \;\; i=n+1, \dots, N \label{variability}
 			\end{align}
 			
 		\end{itemize}
 		
 		The k-means iterative algorithm is  used to classify historical observations  using its formulation in Matlab, \cite{kmeans}.  It returns $k $ clusters $G_1, \dots, G_k$ and their centroids $\mathbf{c_{1}, \dots c_{k}}$; the histograms of these clusters are assumed as the empirical pdfs of the variations with respect to the one-step-ahead irradiance realization. The number of clusters is chosen empirically with the objective of minimizing the variance of each cluster pdf.
 		During real-time operation, the data features vector at time $t$, denoted by $\mathbf{p_{t}} =   ( M_{t}  , V_{t}) $, is calculated.
 		The next step is the calculation of the Euclidean distances between $ \mathbf{p_{t}}  $ and the centroids $\mathbf{c_{l}}$
 		\begin{align}
 		{d_{l}}={\lVert{\mathbf{c_{l}}-\mathbf{p_{t}}}\lVert}^{2}, \;\;   l ={ 1,...,k } 
 		\label{euclidean}
 		\end{align} 
 		which is used as a similarity criterion to select the cluster representative of the future irradiance.
 		We indicate with $ \hat{l} $ the index corresponding to the cluster with minimum distance.  The variance of the cluster pdf is then used as the value for $Q$. In particular:
 		\begin{align}
 		{Q}_{t+1|t} =  \text{Var}(G_{\hat{l}} ) . 
 		\end{align}
 	}}
 	It is worth noting that while doing this, we do the approximation that clusters pdfs are normally distributed. Besides, we note that determining $Q$ requires past irradiance values, whereas previous methods do not. 

\bibliographystyle{IEEEtran}
\bibliography{Bibliography_KF}

\end{document}